\documentclass[11pt,twoside]{article}  % Leave intact
\usepackage{apn3conf}

\begin{document}   % Leave intact

%-----------------------------------------------------------------------
%		            Paper Title 
%-----------------------------------------------------------------------
% Enter the title of the paper.
%
% EXAMPLE: \title{A Breakthrough in Astronomical Software Development}
% 
% If your title is so long as to fill the page header when you print it,
% then please supply a short form as a \titlemark.
%
% EXAMPLE: 
%  \title{Rapid Development for Distributed Computing, with Implications
%         for the Virtual Observatory}
%  \titlemark{Rapid Development for Distributed Computing}
%

\title{Three-Dimensional Adaptive Mesh Refinement Simulations of Point-Symmetric Nebulae}
\titlemark{3D AMR Simulations of Point-Symmetric Nebulae}

%-----------------------------------------------------------------------
%		          Authors of Paper
%-----------------------------------------------------------------------
% Enter the authors followed by their affiliations.  The \author and
% \affil commands may appear multiple times as necessary (see example
% below).  List each author by giving the first name or initials first
% followed by the last name.  Authors with the same affiliations
% should grouped together. 
%
% EXAMPLE: \author{Margaret Meixner\altaffilmark{1}, Letizia Stanghellini,
%			Howard Bond} 
%          \affil{Space Telescope Science Institute, 
%                 3700 San Martin Dr.,  Baltimore, MD 21218}
%          \author{Joel Kastner}
%          \affil{Rochester Institute of Technology}
%
%          \altaffiltext{1}{Astronomy Department, UIUC}
%
% In this example, the first three authors, "Meixner", "Stanghellini", and
% "Bond" are affiliated with "STScI".  "Meixner" has an alternate 
% affiliation with the "Astronomy Department at UIUC".  The fourth author,
% "Kastner", is affiliated with "Rochester Institute of Technology"

\author{Erik-Jan Rijkhorst, Vincent Icke, Garrelt Mellema}
\affil{Sterrewacht Leiden, P.O. Box 9513, 2300 RA, Leiden, The Netherlands}

%-----------------------------------------------------------------------
%			 Contact Information
%-----------------------------------------------------------------------
% This information will not appear in the paper but will be used by
% the editors in case you need to be contacted concerning your
% submission.  Enter your name as the contact along with your email
% address.
% 
% EXAMPLE:  \contact{Dennis Crabtree}
%           \email{crabtree@cfht.hawaii.edu}
%

\contact{Erik-Jan Rijkhorst}
\email{rijkhorst@strw.leidenuniv.nl}

%-----------------------------------------------------------------------
%		      Author Index Specification
%-----------------------------------------------------------------------
% Specify how each author name should appear in the author index.  The 
% \paindex{ } should be used to indicate the primary author, and the
% \aindex for all other co-authors.  You MUST use the following
% syntax: 
%
% SYNTAX:  \aindex{LASTNAME, F. M.}
% 
% where F is the first initial and M is the second initial (if
% used).  This guarantees that authors that appear in multiple papers
% will appear only once in the author index.  
%
% EXAMPLE: \paindex{Crabtree, D.}
%          \aindex{Manset, N.}        
%          \aindex{Veillet, C.}        
%
% NOTE: this information is also used to build the author list that
% appears in the table of contents.  Authors will be listed in the order
% of the \paindex and \aindex commmands.
%

\paindex{Rijkhorst, E.-J.}
\aindex{Icke, V.}
\aindex{Mellema, G.}

%-----------------------------------------------------------------------
%		      Author list for page header	
%-----------------------------------------------------------------------
% Please supply a list of author last names for the page header. in
% one of these formats:
%
% EXAMPLES:
% \authormark{LASTNAME}
% \authormark{LASTNAME1 \& LASTNAME2}
% \authormark{LASTNAME1, LASTNAME2, ... \& LASTNAMEn}
% \authormark{LASTNAME et al.}
%
% Use the "et al." form in the case of seven or more authors, or if
% the preferred form is too long to fit in the header.

\authormark{Rijkhorst, Icke, \& Mellema}

%-----------------------------------------------------------------------
%			Subject Index keywords
%-----------------------------------------------------------------------
% Enter up to 6 keywords describing your paper.  These will NOT be
% printed as part of your paper; however, they will be used to
% generate an object index and a subject index for the proceedings.  
% There is no standard list,  however, individual object names are
% encouraged and one or two word descriptions of the topics (e.g.MHD, 
% ionized gas) are useful. 
%
% EXAMPLE:  \keywords{NGC 7027, AFGL 2688, HD 161796, binary stars,
%                      dust,  molecular gas}
%


%-----------------------------------------------------------------------
%			       Abstract
%-----------------------------------------------------------------------
% Type abstract in the space below.  Consult the User Guide and Latex
% Information file for a list of supported macros (e.g. for typesetting 
% special symbols). Do not leave a blank line between \begin{abstract} 
% and the start of your text.

\begin{abstract}          % Leave intact
Previous analytical and numerical work shows that the {\it generalized interacting stellar winds} model can explain the observed bipolar shapes of planetary nebulae very well.
However, many circumstellar nebulae have a multipolar or point-symmetric shape.
With two-dimensional calculations, Icke showed that these seemingly enigmatic forms can be easily reproduced by a two-wind model in which the confining disk is warped, as is expected to occur in irradiated disks.
In this contribution we present the extension to fully three-dimensional adaptive mesh refinement simulations of such an interaction.
\end{abstract}

%-----------------------------------------------------------------------
%			      Main Body
%-----------------------------------------------------------------------

\section{Introduction}
The interaction between a slow disk-shaped inner AGB nebula and a fast stellar wind as a mechanism to produce bipolar bubbles is by now well established. 
In this {\it generalized interacting stellar winds} model (see Balick \& Frank 2002 for a review) the shape of the dense gas around the star is {\it assumed} to be a disk or a toroid.
Analytical (Icke 1988; Icke et al. 1989) and numerical (e.g. Soker \& Livio 1989; Icke et al. 1992; Mellema et al. 1991) work shows that this mechanism works very well.

However, many circumstellar nebulae have a multipolar or point-symmetric shape (Schwarz 1993; Sahai \& Trauger 1998).
Icke (2003, this volume) proposed that these nebulae are formed in a wind-disk interaction where {\it the disk confining the fast wind is warped}.
One can produce such a warp around a single star, from the combined effects of irradiation and cooling (e.g. Pringle 1996; Maloney et al. 1996).

Whereas Icke's computations were restricted to a two-dimensional proof-of-principle, we now present a first series of fully three-dimensional hydrodynamic computations of such a wind-disk interaction.
%
%---------------------------------------------------------------------------
%
\section{Radiation driven warping}
When an accretion disk is subject to external torques, it may become unstable to warping and when irradiated by a sufficiently luminous central star even an initially flat disk will warp (e.g. Pringle 1996; Maloney et al. 1996).
For this mechanism to work, it is essential that the disk is optically thick for the stellar radiation {\it and} for its own cooling flux.
This restricts the disks in our model to a specific subclass with relatively high density and low temperature.

For a PN the luminosity of the central star alone is sufficiently high to induce a radiation driven warp.
Following Pringle (1997), an expression for the radius $R_{crit}$ beyond which the disk is unstable to radiation driven warping is found:
\begin{equation}
\label{eq:critRadius}
  R_{crit} = (2\pi/A)^2 \; ,
\end{equation}
with the contant $A$ defined by $A^2 \equiv 1/4\,c^{-2} G^{-1} M_*^{-1} L_*^{2} \eta^{-2} \dot{M}_{acc}^{-2}$.
Here $c$ is the speed of light, $G$ the gravitational constant, $\eta\equiv\nu_2/\nu_1$ is the ratio of the azimuthal to the radial viscosity, $M_*$ is the mass and $L_*$ the luminosity of the central star and $\dot{M}_{acc}$ is the disk's accretion rate.
We assumed a surface density $\Sigma_d=\dot{M}_{acc}/(3\pi\nu_1)$ (e.g. Pringle 1981).

In a cartesian coordinate system, the warped disk surface is given by
\begin{equation}
\label{eq:diskSurface}
  \mathrm{\bf x}(R,\phi)=
  R\left(
    \begin{array}{rcl}
       \cos\phi\sin\gamma & + & \sin\phi\cos\gamma\cos\beta\\
      -\cos\phi\cos\gamma & + & \sin\phi\sin\gamma\cos\beta\\
                          & - & \sin\phi\sin\beta
  \end{array}
  \right) \;,
\end{equation}
with local disk tilt angle $\beta(R,\phi)$, and orientation angle of the line of nodes $\gamma(R,\phi)$.
Here, $R$ and $\phi$ are the non-orthogonal radial and azimuthal coordinates, respectively, pointing to the surface of the disk (Pringle 1996).
In our model calculations we adopt the case of a steady precessing disk with no growth and zero torque at the origin.
This gives $\gamma=A\sqrt{R}$ and $\beta=\sin\gamma/\gamma$ (Maloney et al. 1996) in the precessing frame.
%
%---------------------------------------------------------------------------
%
\begin{figure}
\plotone{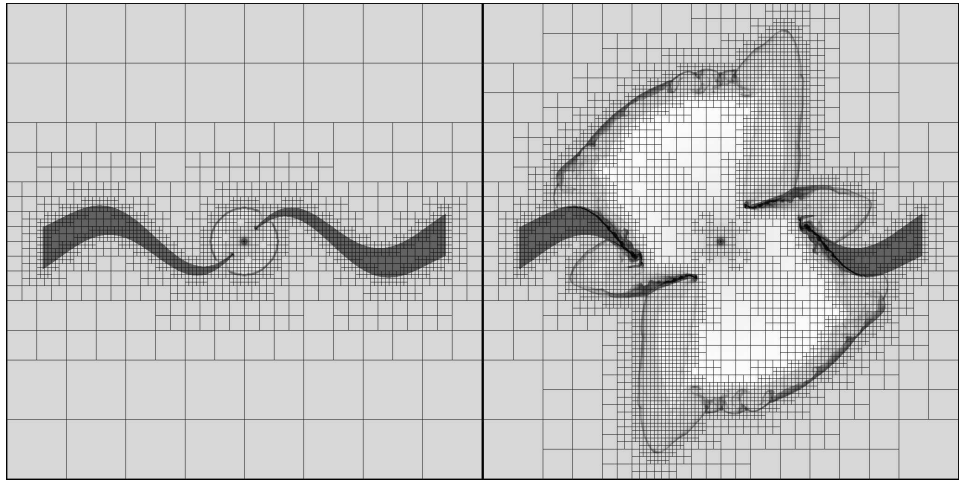}
\caption{2D example of a wind-disk interaction showing the evolving AMR grid structure superimposed on a grayscale plot of the logarithm of the density. Every square represents a grid of 8x8 cells. Five different levels of refinement are visible. The effective resolution of this simulation is 1024x1024 cells.}
\label{fig:grid}
\end{figure}
%
%---------------------------------------------------------------------------
%
\section{3D AMR Simulations}
We used the three-dimensional hydrocode {\em Flash} (Fryxell et al. 2000) to model the interaction between a spherical wind and a warped disk.
This parallelized code implements block-structured adaptive mesh refinement (AMR) initially developed by Berger \& Oliger (1984) and a PPM type hydrosolver (Woodward \& Colella 1984).

To construct the warped disk, Eq.(\ref{eq:critRadius}) was combined with a constant `wedge angle' $\theta_d$ and a proper value for $A$, i.e. $R_d$ was taken to be a few times $R_{crit}\simeq 1\:\mathrm{AU}$, see Eq. (\ref{eq:diskSurface}).
This disk was given a constant density which, through the density contrast $\chi$, resulted in a value for the environment density $n_e$.
The spherical wind was implemented as an inner boundary condition and given a $1/r^2$ density profile and a constant wind velocity $v_w$.
The pressure was calculated from an equation of state with a constant Poisson index $\gamma = 1.1$, resulting in a highly compressible, momentum-driven flow.

To test our code we ran a number of two-dimensional simulations, an example of which is shown in Fig.\ref{fig:grid}.
\begin{figure}
\plotone{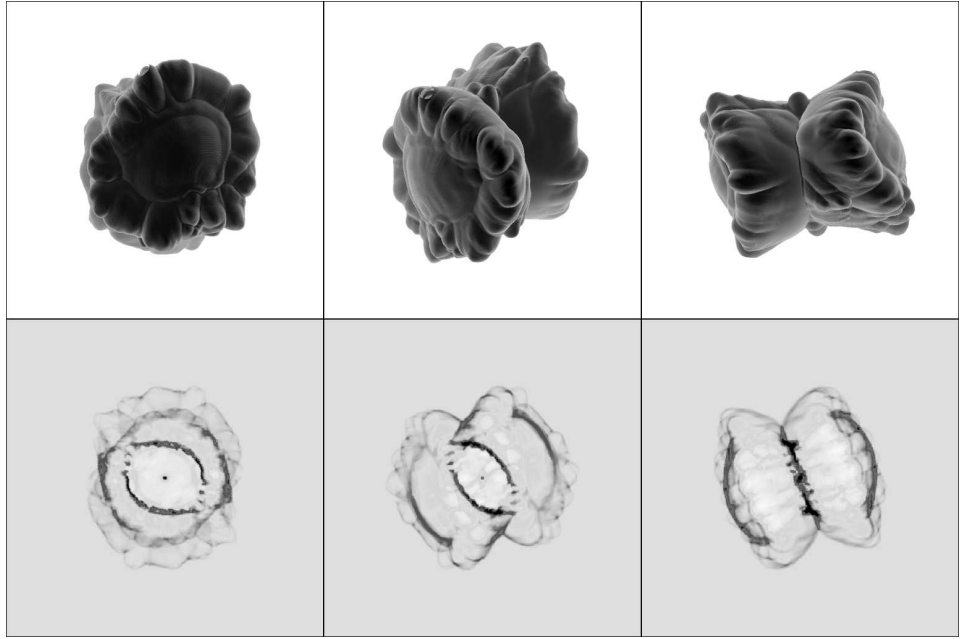}
\caption{Isosurfaces of the density at the end of the wind-disk simulation as seen at different angles ({\it top row}) and the corresponding synthesized $H\alpha$ images ({\it bottom row}).}
\label{fig:iso_proj}
\end{figure}
After this we ran wind-disk simulations in three dimensions on a cartesian grid with an effective resolution of up to $512^3$ cells using five levels of refinement.
We used the following parameters: $n_e = 10^{9}\,\mathrm{cm}^{-3}$, $\chi=100$, $\theta_d=5^o$, and $v_w=200\:\mathrm{km}\,\mathrm{s}^{-1}$.
In Fig.\ref{fig:iso_proj} we present a visualization of the three-dimensional shape of the swept up shell through isosurfaces at different viewing angles.
We also derived the corresponding synthesized $H\alpha$ images by projecting the three-dimensional data cube onto the plane of the sky.
For this, we simply integrated the density squared along the line of sight and used this as a rough estimate for the emission.
The images show that the interaction of a spherical stellar wind with a warped disk results in a wide variety of point-symmetric shapes.
Movies of these simulations can be found at \htmladdURL{http://www.strw.leidenuniv.nl/AstroHydro3D/}.
%
%---------------------------------------------------------------------------
%
\acknowledgements
V.I. expresses his gratitude to Raghvendra Sahai and Hugo Schwarz for lively discussions that were the primary cause for taking up this subject.

The research of G.M. has been made possible by a fellowship of the Royal Netherlands Academy of Arts and Sciences.

The software used in this work was in part developed by the DOE-supported ASCI/Alliance Center for Astrophysical Thermonuclear Flashes at the University of Chicago.

Our work was sponsored by the National Computing Foundation (NCF) for the use of supercomputer facilities, with financial support from the Netherlands Organization for Scientific Research (NWO), under grant number $614.021.016$.
%
%-----------------------------------------------------------------------
%			      References
%-----------------------------------------------------------------------
%

% Do not place any material after the references section

\end{document}